\documentclass[pra,aps,floats,letterpaper,floatfix,footinbib,preprintnumbers,twocolumn]{revtex4}
\usepackage{graphicx}
\usepackage{natbib}
\begin{document}
%
\title{Magnetic field imaging with atomic Rb vapor}

\author{Eugeniy E. Mikhailov}
\author{I. Novikova}
\affiliation{Department of Physics, The College of William~$\&$~Mary, Williamsburg, VA 23187}
\author{M.~D. Havey}
\affiliation{Department of Physics, Old Dominion University,
Norfolk, VA 23529}
\author{F.~A. Narducci}
\affiliation{EO Sensors Division, Naval Air Systems Command, Patuxent River, MD 20670}

\begin{abstract}

We demonstrate the possibility of dynamic imaging of magnetic fields using electromagnetically induced
transparency in an atomic gas. As an experimental demonstration we employ an atomic Rb gas confined in a glass
cell to image the transverse magnetic field created by a long straight wire.  In this arrangement, which clearly
reveals the essential effect, the field of view is about $2\times2$~mm$^2$ and the field detection uncertainty
is 0.14~mG per $10~\mu$m x $10~\mu$m image pixel.

\end{abstract}

\pacs{270.1670, 020.1670, 020.3690, 020.7490}

\date{\today}

\maketitle

Efforts to develop measurement techniques for precision magnetometry have a long and successful history in
atomic physics \cite{Ref:BudkerRomalisReview,wynands02APB,romalisNature03,budkerRSI06,stamperkernPRL07}. Some
recent advances in development of compact atomic magnetometers take advantage of the unique properties of the
electromagnetically induced transparency (EIT) effect and the existence of coherently produced optical dark
states, which can be extremely sensitive to external fields~\cite{Ref:EITReviewArticle}. Here, we propose the
idea of a three axis magnetometer for dynamic imaging of magnetic field gradients based on EIT using state  of
the art imaging technologies. We also demonstrate the essential idea with a one axis (two dimensional)
prototype. As described later, a fully implemented three dimensional magnetic gradiometer promises both high
precision and useful visualization tools for both static and dynamic magnetically complex environments. The
potential applications of this technique range from essentially fundamental (e.g. precision searches for
magnetic monopoles ~\cite{Ref:DiracMonopole}) to quite applied, such as, for example, precise monitoring of the
magnetic environment in medical applications or quantum information
experiments~\cite{Ref:LukinColloquium,KimbleNature08}.

The proposed approach relies on the spectrally ultranarrow transmission resonances observed under EIT conditions
in an atomic gas. These resonances are associated with the coherent non-interacting (dark) states in a
three-level $\Lambda$ system formed under the combined action of two electromagnetic fields in two-photon Raman
resonance with two long-lived metastable states of an atom. The width of the EIT resonances is determined by the
intensities of the laser field and the lifetime of the ground-state coherence, and can be as small as a few
Hz~\cite{budker98prl,erhard01pra}. Such spectral sensitivity makes this effect very promising for precision
metrology. In particular, the spectral location of the transparency resonance depends on the external magnetic
field when the $\Lambda$-system is based on the magnetic field-sensitive Zeeman sublevels. Moreover, if the
magnetic field is spatially varying transversely in a plane orthogonal to the probe beam's $\textbf{k}$ vector,
the resonances will occur at different spatial locations for different two-photon resonance detunings. Thus, the
spatial distribution of transmitted intensity will display a transverse image of the spatial regions where the
two photon resonance condition is satisfied. Recording a series of such images for various two-photon
frequencies can create a spatial map of the transverse magnetic field. Combining such measurements made in three
orthogonal directions (formed from, for example, three independent arrangements) allows six elements of the
$\nabla\textbf{B}$ tensor to be measured. According to the source free Maxwell's Equations, there are four
constraints on the nine components of $\nabla\textbf{B}$, implying a desirable unit redundancy in each set of
six measurements.

\begin{figure} 
\includegraphics[width=0.8\columnwidth]{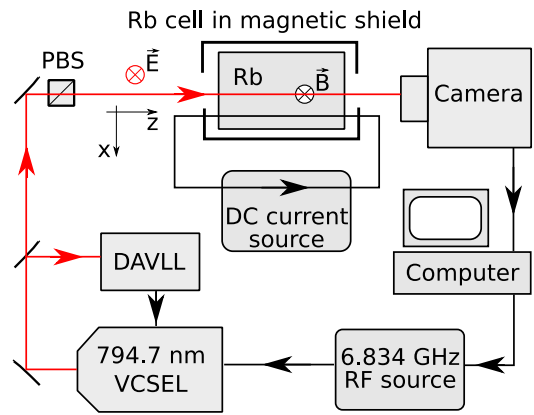}
\caption{\label{fig:setup_gradiometer} A schematic diagram of the experimental configuration.}
\end{figure} 

In this Letter we illustrate the concept by imaging the magnetic field produced by a long straight wire carrying
a steady current $I$ running along the length of a Rb vapor cell. Since in our experiment we are able to detect
the light transmission through the cell only in one direction, we took extreme care to make the wire as coaxial
as possible with the laser beam to remove any variation of the magnetic field along the optical axis. The aim is
then to measure the resulting transverse variation of the magnetic field across the laser beam by imaging the
transverse variations of the transmitted light intensity as a function of the two-photon detuning.

A schematic of the experimental setup is shown in Fig.~\ref{fig:setup_gradiometer}. We use a vertical-cavity
surface emitting laser (VCSEL) directly phase-modulated at 6.834~GHz (a ${}^{87}$Rb hyperfine frequency), and
use the fundamental (carrier) laser frequency and one of the side bands to produce the two optical fields
required for EIT observation. The laser beam with total power 200~$\mu$W and a slightly elliptical Gaussian
profile [waist sizes 1.8~mm and 1.4~mm full width half maximum (FWHM)], was directed into the cylindrical Pyrex
cell (with length 75~mm and diameter 22~mm) containing isotopically enriched $^{87}$Rb vapor and 15~Torr of Ne
buffer gas. The cell was mounted inside a three-layer magnetic shielding to reduce stray laboratory magnetic
fields with the suppression factor of at least $1000$, and its temperature was actively stabilized at 322~K by
regulating the current through a bifiler heater wire wrapped around the innermost shield.

\begin{figure} 
\includegraphics[width=0.7\columnwidth]{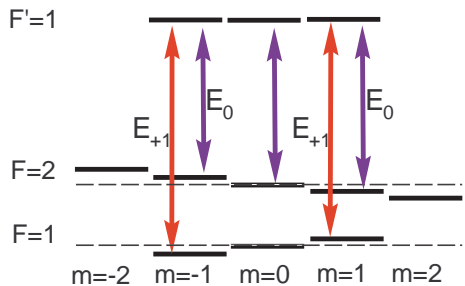}
\caption{Energy level diagram illustrating the Zeeman transitions essential to the EIT $\Lambda$ schemes formed
by the carrier $\mathrm{E_0}$ and the first modulation sideband $\mathrm{E_{+1}}$.} \label{Fig:ZeemanLevels}
\end{figure} 

The magnetic field producing wire was mounted parallel to the laser beam at a distance $\rho_0=20.1 \pm 0.2$~mm,
and connected to a low-noise current supply operating at $I=438\pm 1$~mA. Then the magnetic field outside the
wire produced by this current is orthogonal to the light propagation direction, and can be estimated as
(neglecting the effects of the magnetic shielding):
\begin{equation} \label{BS1}
B(\rho) = \frac{\mu_{o} I}{2\pi \rho}
\end{equation}
\noindent where $\rho$ is the distance from the center of the wire, and $\mu_0$ is permeability of free space.
If we let the laser beam's wave vector and the direction of current flow define the $z$-direction, the curves of
constant magnetic field $B$ are then circles in the x-y plane and centered on the wire. For a given two-photon
detuning the EIT resonance conditions are obeyed only for a given value of the magnetic field, and thus for a
large laser beam we in general expect to observe a corresponding ``bright'' circular arc in the transmission
spatial profile, with the radius of the arc depending on the set two-photon detuning. In our experimental
arrangements, however, the laser beam diameter is small compare to the distance to the wire $\rho_0$, so the
magnetic field across the laser beam is mainly along the $y$-axis and changes linearly with small displacement
$\delta x$:
\begin{equation} \label{BS2}
B(\rho) = \frac{\mu_{o} I}{2\pi \rho_0}\left(1-\frac{\delta x}{\rho_0} \right)
\end{equation}

For the measurements, the laser frequency was tuned such that the stronger optical field (unmodulated carrier)
was resonant with the $5~S_{1/2}$ $F=2 \rightarrow 5~P_{1/2}$ $F'=1$ transition of ${}^{87}$Rb, while the
frequency of the $+1$ modulation sideband matched the $5~S_{1/2}$ $F=1 \rightarrow 5~P_{1/2}$ $F'=1$ transition.
The laser output was linearly polarized. In this configuration there are two possible $\Lambda$-systems for
$F=1,2; m_F=\pm 1 \rightarrow F'=1; m_{F'}=\pm 1$ that produce EIT resonances at two-photon detunings $\Delta
\nu = \Delta_{HFS} \pm 2 g_m \mathrm{B}$ (shown in Fig.~\ref{Fig:ZeemanLevels}), where $g_m =0.7$~MHz/G is the
gyromagnetic ratio for $^{87}$Rb~\cite{note1}. We also adjusted the power of the VCSEL modulation signal to
cancel the light shift of the EIT resonances, and to avoid the effects of non-uniform spatial distribution of
the laser beam on the measured resonance position~\cite{linlinpreprint}. For our experimental conditions such
cancelation occurred for a sideband/carrier intensity ratio of 0.75. Also, we analyzed only the area of the
image where the intensity of the beam exceeds half of its maximum value.

\begin{figure*} 
\includegraphics[width=\textwidth]{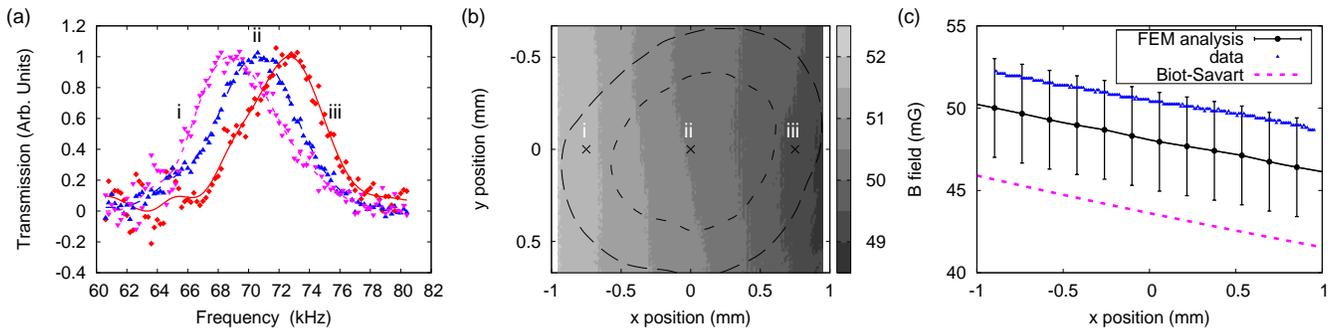}
\caption{\emph{(a)} Transmission through the cell detected at three different camera pixels [marked in
\emph{(b)}] as functions of the two-photon detuning. Solid lines show low-pass filtered data used to determine
the maximum transmission position. \emph{(b)} Map of the magnetic field strength. Dashed lines mark the 75\% and
50\% levels of the maximum laser beam intensity. \emph{(c)} Experimentally measured magnetic field slice of the
map \emph{(b)} along $y=0$ as a function of $x$ position, and the theoretically calculated magnetic fields using
the Biot-Savart law [Eq.(\ref{BS1})] and a finite element analysis. The error bars in the FEM curve represent
the variation in the calculated magnetic field due to $\pm 1$~mm uncertainty in the magnetic shielding position
with respect to the wire and the laser beam. \label{fig:exp_results} }
\end{figure*}

To map the spatial variation of the magnetic field, we stepped the VCSEL modulations frequency in a 20~kHz range
around one of the EIT resonances in 200~Hz increments, taking images of the transmitted laser beam profile using
a digital camera (Unibrain Fire-i 511b). The dominant source of noise in our experiment was due to the
phase-to-amplitude noise conversion of the VCSEL large phase noise (the linewidth of the laser was $\approx$
100~MHz) by the absorption in the cell. To reduce the detection noise of the image,  we recorded and averaged
200 sequential images grabbed from the camera, which operated at a 30 frame per second rate for any particular
two photon detuning. Even then the detection noise is larger than the digital resolution of the 12 bit
analog-to-digital converter of the camera. For each pixel of the image we plotted the intensity as a function of
the two-photon detuning, as shown in Fig.~\ref{fig:exp_results}(a). Then we located the
position of the EIT maximum (and correspondingly the value of the local magnetic field). 
After analyzing the transmission for every pixel  we obtained a spatial map of magnetic field, shown in
Fig.~\ref{fig:exp_results}(b). The average variance between two sequential runs allowed us to put an upper limit
on the experimental uncertainty  of the
magnetic field measurements $\Delta B = 0.14$~mG for every $10~\mu$m pixel. 

%

We compare the measured magnetic field with numerical calculation of the magnetic  field  generated by a long
straight wire inside a cylindrical  magnetic shield.  Since  the  inner  shield modifies the magnetic  field
from the simple Biot-Savart law given by Eq.~\ref{BS1}, we used a finite element (FEM) analysis program to
calculate the expected magnetic field distribution inside the laser beam~\cite{FEMnote}, as shown in
Fig.~\ref{fig:exp_results}(c). The rather large uncertainty of our FEM model is governed by possible variations
in each shield's diameter, thickness, and imperfect cylindrical shapes, as well as the uncertainty in the
location of the wire inside of the shielding. For instance, Fig.~\ref{fig:exp_results}(c) illustrates that
displacement of the shielding position as small as 1~mm is sufficient to match the FEM simulations with the
experimental measurements.




The demonstrated sensitivity of our apparatus ($\approx 14\mu G/\mu m$) is limited by the transmitted intensity
fluctuations of the broadband VCSEL output. We expect much better sensitivity for narrow-band electromagnetic
fields, which should dramatically reduce the amount of amplitude noise in the output. A good candidate for such
a source, for example, could be a narrow band external cavity diode laser followed by an external phase
modulator.

We can estimate the precision of the magnetic field measurements $\Delta B$ due to the uncertainty in two-photon
detuning frequency corresponding to the maximum transmission for a given signal-to noise ratio $S/N$
~\cite{Fourier_book}:
\begin{equation}
\Delta B \simeq \frac{1}{2 g_m} \frac{\gamma}{S/N \sqrt{n}}.
\end{equation}
In this expression we assume that the EIT resonance has Lorentzian shape with a known FWHM $\gamma$, and $n$ is
the number of sampling points per resonance width. The optimal frequency span in this case should exceed, or be
comparable with, the value $\gamma$, with $n>5-10$ for each recorded trace to avoid numerical errors in fitting
related to discretization artifacts.

%
%

We note that the ultimate relative sensitivity of our method should be similar to that of atomic clocks, based
on the same operational principle~\cite{vanier05apb}. In particular, miniature atomic clocks demonstrate a
fractional stability on the order of $10^{-11}-10^{-12}/\sqrt{\mathrm{Hz}}$. However, in practice the
performance of the magnetic field imager may be limited by other factors, such as digital noise or the limited
optical sensitivity per pixel of a charge coupled camera. These, and other factors including the essential role
of longitudinally varying magnetic fields, are currently under investigation.

In conclusion, we propose a new method of dynamic imaging of a magnetic field by detecting spatial and temporal
variations in transmission of light through an atomic sample under conditions of electromagnetically induced
transparency. In general, this method can be applied to measurements of three-dimensional magnetic field maps.
To demonstrate the basic operational principle, we reconstructed the spatial distribution of the transverse
magnetic field created by a current-carrying long straight wire, running along an atomic Rb vapor cell, with the
precision of $14\mu G/\mu m$ inside the $1.8\times1.4$~mm$^2$ laser beam cross-section.
Our prototype apparatus can be readily scaled up for either a
larger or smaller detection area by adding a zooming telescope in
front of the camera (although the total power of the laser would
have to be reoptimized). Also, since our prototype experiment was
performed using a VCSEL, we expect that this method can be used
with recently developed chip-scale systems for miniature atomic
clocks and magnetometers~\cite{NISTmagnetometer}.

The authors thank D. Malyarenko and J.~P. Davis for useful discussions, and S. Aubin for his help  with the
image acquisition code. This research was supported by NSF grants PHY-0758010, PHY-0654226, Jeffress Research
grant J-847, ONR grants N0001409WX20679 and In-House Laboratory Innovative Research (ILIR).







\begin{thebibliography}{99}
\bibitem{note1}
For arbitrary orientation of the laser polarization with respect of the magnetic field vector various $\Lambda$
system will be formed by both $\pi$ and $\sigma_\pm$ transitions, resulting in EIT resonances at $0, \pm g_mB$
and $\pm 2 g_mB$ two-photon detunings. Careful analysis of relative amplitudes of these resonances may allow
measurements of not only the magnitude but also the direction of the magnetic field.
\end{thebibliography}
\end{document}